\documentclass[preprint,superscriptaddress, amsmath,amssymb, aps]{revtex4-1}

\usepackage{graphicx}
\usepackage{dcolumn}
\usepackage{bm}
\usepackage{xcolor}
\usepackage{colortbl,booktabs}
\usepackage{float}
\usepackage{slashed}
\usepackage{subfigure}
\usepackage[colorlinks=true,linkcolor=red,citecolor=blue]{hyperref}
\allowdisplaybreaks 
\begin{document}

\title{$D^*D\rho$ and $B^*B\rho$ strong couplings in light-cone sum rules}

\author{Chao Wang}\email{chaowang@nankai.edu.cn}
\affiliation{\it \small School of physics, Nankai University, Weijin Road 94, 300071 Tianjin, China}
\author{Hua-Dong Li}\email{lihd@ihep.ac.cn}
\affiliation{\it \small Institute of High Energy Physics, CAS, P.O. Box 918, Beijing 100049, China}
\affiliation{\it \small School of Physics, University of Chinese Academy of Sciences, Beijing 100049, China}

\begin{abstract}
We present an improved calculation of the strong coupling constants $g_{D^*D\rho}$ and $g_{B^*B\rho}$ in light-cone sum rules  including the one-loop QCD corrections of leading power with $\rho$ meson distribution amplitudes. We further compute the subleading-power corrections  from two-particle and three-particle higher-twist contributions at leading order up to twist-4 accuracy.   The next-to leading order corrections to leading power contribution  offset the subleading-power corrections to certain extend numerically, and our numerical results are consistent with previous works from sum rules. The comparisons between our results and the existing model-dependent estimations are also made.
\end{abstract}

\maketitle

\section{Introduction}
This paper aims to give a more precise determination of $D^*D\rho$ coupling and $B^*B\rho$ coupling in the framework of light-cone sum rules (LCSR), which describe the low-energy interaction among heavy mesons and light mesons, are of great importance to understand the QCD long-distance dynamics.  The coupling is a fundamental parameter of the effective Lagrangian of heavy meson chiral perturbative theory (HM$\chi$PT) \cite{Casalbuoni:1996pg,Yan:1992gz,Wise:1992hn,Burdman:1992gh}, which plays an important role in the study of heavy meson physics. Phenomenologically, it describes the strength of the final state interactions \cite{Cheng:2004ru} which are important in the generation of the strong phase within B decays \cite{Cheng:2016shb,Virto:2016fbw}. Moreover, the coupling relates the pole residue of $D(B)$ to $\rho$ form factors at large momentum transfer by the dispersion relation, which is helpful for us to gain a better understanding on the behavior of form factors.

Various theoretical approaches to determine the coupling have been suggested. Firstly, with the $D(B)$ to $\rho$ form factors obtained at certain region of the momentum transfer from LCSR \cite{Ball:2004rg,Straub:2015ica,Gao:2019lta} and the lattice QCD  \cite{Bowler:2004zb},  the corresponding pole residue which relates the coupling can be extracted with appropriate extrapolation for the form factors. The simplest way to do such extrapolation is the vector meson dominance (VMD) hypothesis which neglects the continuum spectral. Except for the VMD approximation, there are also some other modified parameterizations for the form factors that have been proposed in \cite{Becirevic:1999kt,Ball:2004ye,Boyd:1994tt,Bourrely:2008za}. Apart from above method, the coupling can also be estimated  from HM$\chi$PT with VMD approximation as done in \cite{Casalbuoni:1992gi,Casalbuoni:1992dx,Casalbuoni:1993nh,Isola:2003fh}.

Another strategy to obtain the strong coupling is calculation from first principles of QCD. We study the strong coupling constants $g_{D^*D\rho}$ and $g_{B^*B\rho}$ in LCSR by using double dispersion relation. The LCSR  was proposed in  \cite{Balitsky:1989ry,Braun:1988qv,Chernyak:1990ag} based on the light-cone operator-product-expansion (OPE) relative to the conventional QCD sum rules (QCDSR) method.  There have been several works regarding the couplings $g_{D^*D\rho}$ and $g_{B^*B\rho}$, starting from \cite{Aliev:1996xb} with inclusion of two-particle $\rho$ DAs corrections up to twist-3 at leading order (LO). Years later,  \cite{Li:2002pp,Li:2007dv,Wang:2007mc,Wang:2007zm} improved the formal calculation by considering the two-particle twist-4 corrections \cite{Ball:1998ff} at LO. Meanwhile, $g_{D^*D\rho}$ is also calculated with three-point QCDSR by taking into account the dimension-5 quark-gluon condensate corrections under flavor $SU(3)$ symmetry \cite{Khosravi:2014rwa}. Concerning previous works, estimations for the couplings exhibit widespread value. In the LCSR works, their results \cite{Li:2002pp,Li:2007dv,Wang:2007mc,Wang:2007zm} are smaller than other estimations, so we need to proceed a more accurate calculation to confirm whether this difference comes from radiative and power corrections.  Moreover, inclusion of NLO corrections also can decrease the scale-dependence. In this work, we give a calculation including $\mathcal{O}(\alpha_s)$ corrections to leading power contribution with the resummation of large logarithm to next-to leading logarithmic accuracy (NLL). For the subleading-power corrections, our results also include the three-particle twist-4 corrections at LO.

The paper is organized as follows: in section \ref{sec-2} we calculate the leading power contributions up to NLO. Procedures for analytic continuation and continuum subtraction are similar to \cite{Belyaev:1994zk,Khodjamirian:1999hb}. Subleading-power corrections including two-particle and three-particle corrections up to twist-4 at LO are calculated in section \ref{sec-3}. Section \ref{sec-4} provides our numerical results and the phenomenological discussion. We will summarize this work in the last section.

\section{The leading power contributions}
\label{sec-2}

\subsection{Hard-collinear factorization at LO in QCD}

The strong coupling constant $g_{H^*H\rho}$ is defined by
\begin{eqnarray}
\langle\, \rho(p,\eta^*)\,H(q)\,|\,H^*(p+q,\varepsilon)\, \rangle= -g_{H^*H\rho}\, \epsilon_{pq\eta\varepsilon}\,,
\label{gdef}
\end{eqnarray}
with $\epsilon_{pq\eta\varepsilon}=\epsilon_{\mu\nu\alpha\beta}\,p^\mu\,q^\nu\,\eta^{*\alpha}\,\varepsilon^\beta$. Here we choose $H^*$ and $H$ stand for $D^{*-}(B^{*0})$ meson and $\bar{D}^{0}(B^+)$ meson respectively, and $\rho$ is $\rho^-$. $\varepsilon_{\mu}$ and $\eta_\nu$ are the polarization vectors of the $H^*$ and $\rho$ mesons respectively. We use the conventions $\epsilon_{0123}=-1$ and $D_{\mu}=\partial_\mu-i\,g_s\,T^a\,A_\mu^a$. The couplings of different charge states are related by isospin symmetry, for instance,
\begin{align}
g_{D^*D\rho}\equiv g_{D^{*-}\bar{D}^0\rho^-}=-\sqrt{2}\,g_{D^{*-}D^-\rho^0}\,.
\end{align}

We construct the following correlation function at the starting point
\begin{align}
\Pi_\mu(p,q)=\int d^4x\,e^{-i(p+q)\cdot x}\langle\,\rho^-(p,\eta^*)\,|\,T\big\{\bar{d}(x)\,\gamma_{\mu \perp}\, Q(x),\,\bar{Q}(0)\,\gamma_5\,u(0)\big\}\,|\,0\rangle\,,
\label{corre-fun}
\end{align}
where $\gamma_{\mu\,\perp}=\gamma_{\mu}-\slashed{n}/2\,\bar{n}_\mu-\slashed{\bar{n}}/2\,n_\mu$, $Q$  is the heavy quark field. We have introduced two light-cone vectors $n_\mu$ and $\bar{n}_\mu$ with $n\cdot n=\bar{n}\cdot \bar{n}=0\,, n\cdot\bar{n}=2$. For the momentum, we choose a coordinate to have $q_{\perp}=0$.  In the frame of LCSR, $\rho$ meson is on the light-cone, and its momentum is chosen as
\begin{align}
p^{\mu}=\frac{n\cdot p}{2}\bar{n}_{\mu}\,.
\label{corre-fun}
\end{align}

On the hadronic level, taking advantage of the following definitions for decay constants
\begin{align}
\langle\,H^*(p+q,\varepsilon^*)\,|\,\bar{d}\,\gamma^\mu \,Q\,|\,0\,\rangle=f_{H^*}\,m_{H^*}\varepsilon^{*\mu}\,,\quad \langle\,0\,|\,\bar{Q}\,\gamma_5\,u\,|\,H(p)\,\rangle=-i\,f_H\,\frac{m_H^2}{m_Q}\,,
\end{align}
the correlation function (\ref{corre-fun}) can be written as
\begin{align}
\Pi_\mu^{\rm had}(p,q) =&~\frac{g_{H^*H\rho}\,f_{H^*}\,f_H}{[m_{H^*}^2-(p+q)^2-i\,0]\,[m_H^{2}-q^2-i\,0]}\,
\frac{m_H^2\,m_{H^*}}{m_Q}\,\epsilon_{\mu pq\eta} \nonumber\\
&+\iint_{\Sigma}\frac{\rho^h(s,s')\,ds\,ds'}{\left[s'-(p+q)^2\right](s-q^2)}+\cdots\,,
\label{hadronic}
\end{align}
where the second term counts the contributions from higher resonances and continuum states. The ellipses denote the terms that vanish after double Borel transformation.

After double Borel transformation we get the hadronic representation of the correlation function
\begin{align}
\Pi_\mu^{\rm had}(p,q)= \frac{f_H\,f_{H^*}\,m_H^2\,m_{H^*}}{m_Q}\,g_{H^*H\rho}\, e^{-\frac{m_H^2+m_{H^*}^2}{M^2}} \epsilon_{\mu pq \eta} +\iint_{\Sigma}\,ds\,ds'\,e^{-\frac{s+s'}{M^2}}\,\rho^h(s,s')\,.
\label{hadronic-borel}
\end{align}
For the boundary of the integral $\Sigma$, we take $s+s'=2\,s_0$ with $s_0$ as the threshold of excited and continuum states.
The Borel parameters associated with $(p+q)^2$ and $q^2$ are quite similar in magnitude, so we set the same value $M^2$.

\begin{figure}[!htb]
\begin{center}
\includegraphics[width=0.40 \columnwidth]{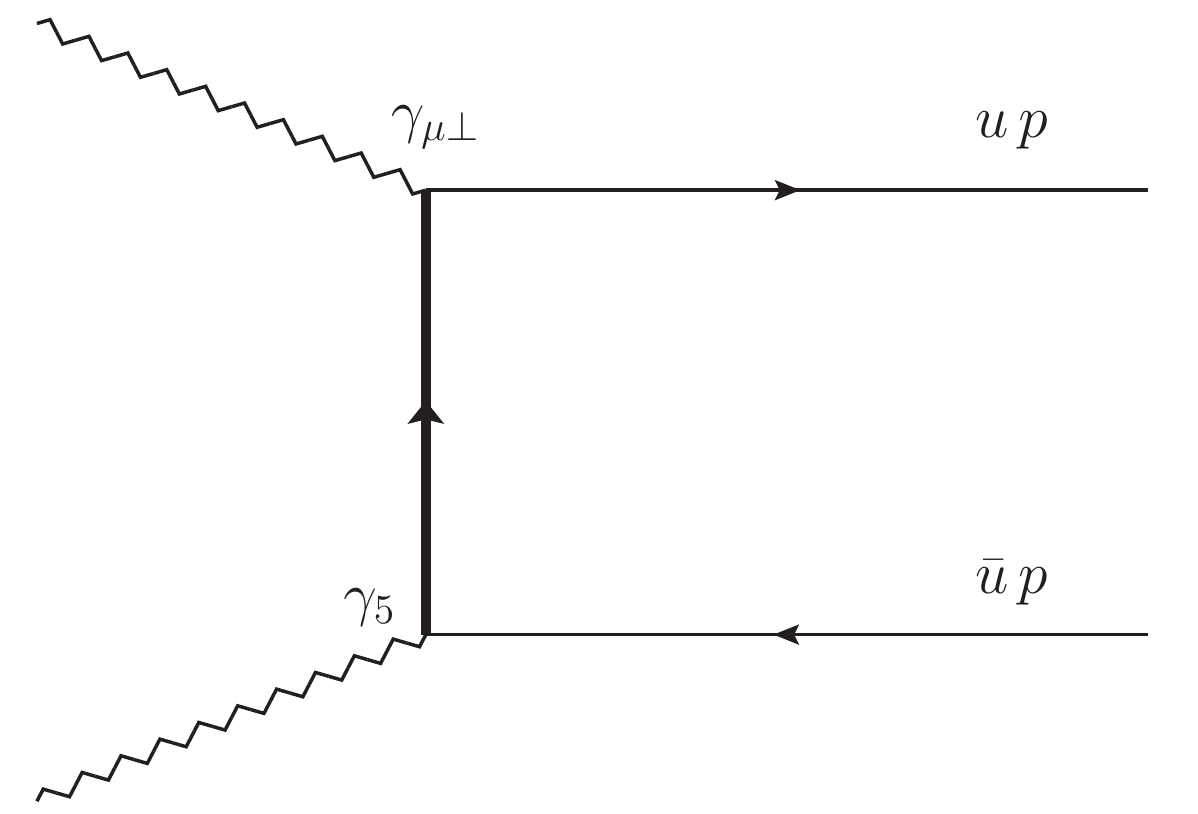} \\
\vspace*{0.1cm}
\caption{Diagrammatical representation of the leading-order (LO) contribution. }
\label{twist-2-LO}
\end{center}
\end{figure}

On the quark level, the leading-twist tree diagram is displayed in Fig. \ref{twist-2-LO}.
The correlation function reads
\begin{align}
\Pi_\mu^{{\rm LT},(0)}(p,\,q)&=-\frac{i}{2}\frac{\bar{n}\cdot q}{u\,q^2+\bar{u}\,(p+q)^2-u\,\bar{u}\,m_\rho^2-m_Q^2}\, \bar{q}(u\,p)\,\gamma_{\mu\,\perp}\,\slashed{n}\,\gamma_5\,q(\bar{u}\,p)\nonumber\\
&=-\frac{i}{2}\frac{\bar{n}\cdot q}{u\,q^2+\bar{u}\,(p+q)^2-m_Q^2}\, \bar{q}(u\,p)\,\gamma_{\mu\perp}\,\slashed{n}\,\gamma_5\,q(\bar{u}\,p)+\mathcal{O}\Big(\frac{\Lambda^2_{\rm QCD}}{m_Q^2}\Big)\,. \label{lpfull}
\end{align}

Using the definition of the leading-twist $\rho$ DA \cite{Ball:1998ff}  in appendix \ref{rhodas}, we get the leading twist tree level factorization formula
\begin{align}
\Pi_{\mu}^{{\rm LT},(0)}(p,\,q)=-f_\rho^T(\mu)\,\epsilon_{\mu p q\eta}\int_0^1 du\,\phi_\perp(u,\mu)\,\frac{1}{u\,q^2+\bar{u}\,(p+q)^2-m_Q^2}\,.
\label{t-2-tree}
\end{align}

The result (\ref{t-2-tree}) can be written in the form of double dispersion relation
\begin{align}
\Pi_{\mu}^{{\rm LT},(0)}(p,\,q)=-f_\rho^T(\mu)\,\epsilon_{\mu p q\eta}\iint ds\,ds'\,\frac{\rho^{LT,(0)}(s,s')}{[s-q^2]\,[s'-(p+q)^2]}\,, \label{quark-tw2lo}
\end{align}
where the double dispersion density $\rho^{LT,(0)}$ is defined as
\begin{align}
\rho^{{\rm LT},(0)}(s,s')=\frac{1}{\pi^2}\,{\rm Im}_{s'}\,{\rm Im}_{s} \int_0^1du\,\frac{\phi_\perp(u,\mu)}{u\,s+\bar{u}\,s'-m_Q^2}\,.
\end{align}

The expression of wave function $\phi_\perp(u,\mu)$ in terms of Gegenbauer polynomials can be written as
\begin{eqnarray}\label{phi}
\phi_\perp(u,\mu)=6\,u\,\bar{u}\,\sum_{n=0}^{\infty}a_n^\perp(\mu)\,C_n^{3/2}(2u-1)=\sum_{k=1}^\infty b_k\,u^k\,,
\end{eqnarray}
where $b_k$ is the function of Gegenbauer moments $a(\mu)$. The spectral density can be obtained as follows \cite{Belyaev:1994zk}
\begin{align}
\rho^{{\rm LT},(0)}(t,v)&=\sum_{k=1}^\infty b_k\,\frac{1}{\pi^2}\,{\rm Im}_{s'}\,{\rm Im}_{s}\int_0^1du\,\frac{u^k}{u\,s+\bar{u}\,s'-m_Q^2} \nonumber\\
&=\sum_{k=1}^\infty b_k\,\frac{(-1)^{k+1}}{k!}\,\frac{1}{2^{k+1}\,t}\,\big(\frac{m_Q^2}{t}-\bar{v}\big)^k\, \delta^{(k)}\big(v-\frac{1}{2}\big)\,\theta(v\,t-m_Q^2)\,.\label{t2loden}
\end{align}
We defined two variables $t=s+s', v=\frac{s}{s+s'}$ in (\ref{t2loden}), which are similar to our work \cite{Li:2020rcg}.

Equating  (\ref{hadronic}) and  (\ref{quark-tw2lo}) and applying double Borel transformation, then subtracting the continuum states by using quark-hadronic duality,  we obtain the leading-twist strong coupling constant at LO
\begin{align}
g^{{\rm LT},(0)}=\frac{m_Q}{f_H\,f_{H^*}\,m_H^2\,m_{H^*}}\,f_\rho^T(\mu)\,\frac{M^2}{2}\, \phi_\perp\Big(\frac{1}{2},\mu\Big)\,
\,\Big[e^{\frac{m_H^2+m_{H^*}^2-2m_Q^2}{M^2}}-e^{\frac{m_H^2+m_{H^*}^2-2s_0}{M^2}} \Big]\,. \label{tree-result}
\end{align}

\subsection{Next to leading order corrections}
\begin{figure}[!htb]
\begin{center}
\includegraphics[width=1.0 \columnwidth]{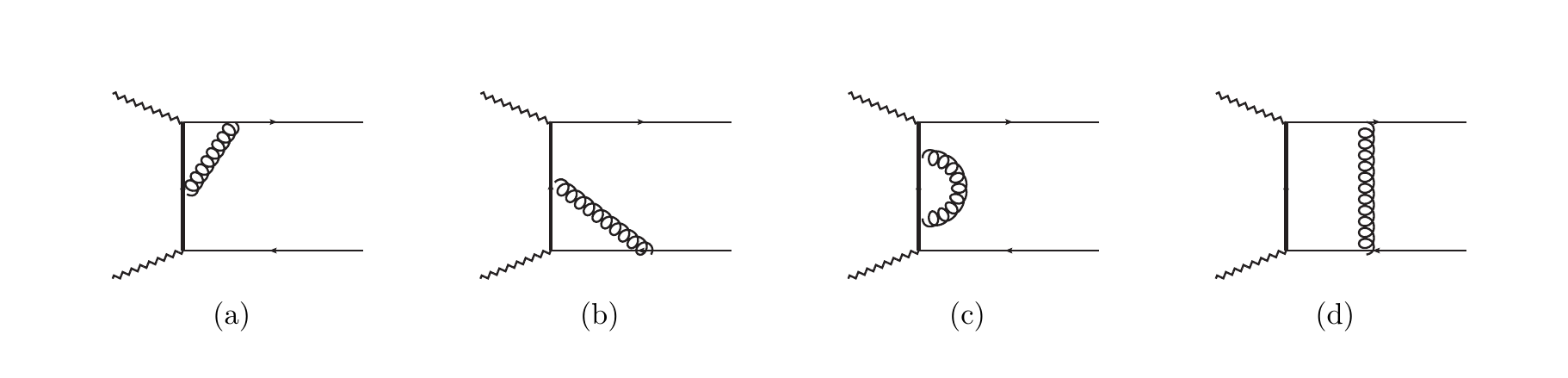} \\
\vspace*{0.1cm}
\caption{NLO QCD corrections of leading-twist contributions.}
\label{twist-2-NLO}
\end{center}
\end{figure}

The one-loop diagrams  are shown in Fig. \ref{twist-2-NLO}. We note that the loop calculations are similar to \cite{Li:2020rcg}. Borrowing the results from it and taking the asymptotic form of wave function in eq (\ref{phi}) as $\phi_\perp(u)=6\,u\,\bar{u}$ for the feasibility of calculation. Subsequently, adopting the similar procedure for the deriving of double dispersion spectral densities, then applying double Borel transformation and continuum subtraction to get the result. We omit the detailed procedures here for simplicity and the explicit operations with final results can be seen from our work \cite{Li:2020rcg}.

Combing the NLO result from \cite{Li:2020rcg} with eq (\ref{tree-result}), we obtain the leading-twist sum rules up to $\mathcal{O}(\alpha_s)$ and the coupling constant reads
\begin{align}
g^{{\rm LT}}=-\frac{m_Q}{f_H\,f_{H^*}\,m_H^2\,m_{H^*}}\, e^{\frac{m_H^2+m_{H^*}^2}{M^2}}\, f_\rho^T(\mu)\,
\Big[-\frac{M^2}{2}\,\phi_\perp\Big(\frac{1}{2},\mu\Big)\,\big( e^{-\frac{2m_Q^2}{M^2}} -e^{-\frac{2s_0}{M^2}}\big) + \mathcal{F}^{{\rm LT},(1)}\Big]\,, \label{twist2nlo}
\end{align}
where
\begin{align}
\mathcal{F}^{{\rm LT},(1)}=\frac{\alpha_s\,C_F}{4\,\pi}\,\Big\{m_Q^2\int_{0}^{2\hat{s}_0-2}d\sigma\, e^{-\frac{\sigma+2}{\hat{M}^2}}\,g(\sigma) +\Delta\,g(\hat{M}^2,m_Q^2)\Big\}\,,
\label{nlo-ftw2}
\end{align}
and
\begin{align}
&g(\sigma)=3\,{\rm Li}_2(-\sigma)+3\,{\rm Li}_2(-\sigma-1)-6\,{\rm Li}_2\big(-\frac{\sigma}{2}\big) -3\,\ln\frac{\sigma}{2}\ln\frac{\sigma+2}{2}+3\,\ln(\sigma+1) \,\ln(\sigma+2) \nonumber\\
&-\frac{6\,(\sigma+1)^2}{(\sigma+2)^3}\,\ln (\sigma+1)
+ \frac{3\,(7\,\sigma^3+50\,\sigma^2+100\,\sigma+64)}{4\,(\sigma+2)^3}\,\ln\frac{\sigma}{2}
+ \frac{3}{2}\,\ln\frac{\sigma+2}{2}\nonumber\\
&+\frac{3\,(11\,\sigma^2+28\,\sigma+24)}{8\,(\sigma+2)^2}-3\,\ln\frac{\mu^2}{m_Q^2} -\frac{9}{4}\,\ln\frac{\nu^2}{\mu^2}-\frac{\pi^2}{4}\,, \nonumber\\
&\Delta\,g(\hat{M}^2,m_Q^2)= 3\,\big(4+3\,\ln\frac{\mu^2}{m_Q^2}\big)\, m_Q^2\,e^{-\frac{2m_Q^2}{M^2}}\,.
\end{align}

For the scale dependence, we use
\begin{align}
\frac{d}{d\ln\mu}m_Q(\mu)=-6\frac{\alpha_sC_F}{4\pi}m_Q(\mu)\,, \quad \frac{d}{d\ln\mu}f_\rho^T(\mu)=-2\frac{\alpha_sC_F}{4\pi}f_\rho^T(\mu)\,,
\end{align}
and we get
\begin{align}
\frac{d}{d\ln\mu}g^{\rm LT}=0+\mathcal{O}(\alpha_s^2)\,,
\end{align}
it's obvious that our result is independent of the factorization scale $\mu$ at one-loop level.

\section{The subleading-power corrections at LO}
\label{sec-3}

In this section, we are going to perform the subleading power corrections to coupling constants, which are involved with two-particle and three-particle corrections up to twist-4. The higher twist (up to twist-4) $Q$-quark propagator in the background field is adopted \cite{Balitsky:1987bk}:
\begin{eqnarray}
&&\langle\,0\,|\,T\{\bar{Q}(x),\,Q(0)\}\,|\,0\,\rangle \nonumber\\
&&\supset i\,g_s\int\frac{d^4k}{(2\pi)^4}\,e^{-i\,k\cdot x}\int_0^1du \,\Big[ {u \, x_{\mu} \over k^2-m_Q^2 } \, G^{\mu \nu}(u\, x) \, \gamma_{\nu}
- {\slashed{k} + m_Q \over 2\,(k^2-m_Q^2)^2} \, G^{\mu \nu}(u\, x) \, \sigma_{\mu \nu}\Big]\,.
\label{charm-propa}
\end{eqnarray}

For the two-particle corrections, inserting the first term of (\ref{charm-propa}) into correlation function (\ref{corre-fun}), and using the DAs of $\rho$ meson, we gain the higher twist two-particle corrections
\begin{align}
&\Pi_{\mu}^{\rm 2P, HT}(p,q) \nonumber\\
=&- \frac{1}{4}\,\epsilon_{\mu pq\eta}\,
\int_0^1du\, \bigg\{ \frac{-2\,m_Q\,f_\rho\,m_\rho\,g_\perp^{(a)}(u)+f_\rho^T(\mu)\,m_\rho^2\,\mathbb{A}_T(u)} {[u\,q^2+\bar{u}\,(p+q)^2-m_Q^2]^2}+ \frac{-2\,m_Q^2\,f_\rho^T(\mu)\,m_\rho^2\,\mathbb{A}_T(u)}{[u\,q^2+\bar{u}\,(p+q)^2-m_Q^2]^3}\bigg\}\,.
\label{p2-higher}
\end{align}
The twist-3 $g_\perp^{(a)}$ corrections are suppressed by $\mathcal{O}(\Lambda_{\rm QCD}/m_Q)$ and twist-4 $\mathbb{A}_T$ corrections are suppressed by $\mathcal{O}(\Lambda_{\rm QCD}^2/m_Q^2)$ comparing with the leading-twist corrections (\ref{quark-tw2lo}). To keep consistent, we should also consider the subleading-power corrections of leading-twist. With respect to (\ref{lpfull}) the correlation function reads
\begin{align}
\Pi_\mu^{\rm LT,NLP}(p,\,q)&=-\frac{i}{2}\,\bar{n}\cdot q\,\frac{u\,\bar{u}\,m_\rho^2}{[u\,q^2+\bar{u}\,(p+q)^2-m_Q^2]^2}\, \bar{q}(u\,p)\,\gamma_{\mu\perp}\,\slashed{n}\,\gamma_5\,q(\bar{u}\,p) \nonumber\\
&=-f_\rho^T\,\epsilon_{\mu pq\eta}\,\int_0^1du\, \phi_\perp(u,\mu)\, \frac{u\,\bar{u}\,m_\rho^2}{[u\,q^2+\bar{u}\,(p+q)^2-m_Q^2]^2}\,.
\label{p2-ltnlp}
\end{align}
Similar to the leading-twist LO case, after applying double Borel transformations and subtracting the continuum states by using quark-hadronic duality, we obtain  the strong coupling constant for two-particle  subleading-power corrections at LO
\begin{align}
g^{{\rm 2P}}=&-\frac{m_Q}{f_H\,f_{H^*}\,m_H^2\,m_{H^*}}\,e^{\frac{m_H^2+m_{H^*}^2-2m_Q^2}{M^2}}\,\frac{m_\rho}{2} \, \Big[\frac{m_\rho}{2}\,f_\rho^T(\mu)\,\phi_\perp\Big(\frac{1}{2},\,\mu\Big) \nonumber\\
&-m_Q\,f_\rho\,g_\perp^{(a)}\Big(\frac{1}{2},\,\mu\Big) +m_\rho\,f_\rho^T(\mu)\,\Big(\frac{1}{2}+\frac{m_Q^2}{M^2}\Big)\, \mathbb{A}_T\Big(\frac{1}{2},\,\mu\Big)\Big]\,.
\label{ht2p-result}
\end{align}

Considering  three-particle corrections at tree level, the correlation function of three-particle $q\bar{q}g$ corrections can be written as
\begin{align}
\Pi_{\mu}^{{\rm3P}}(p,q)
=&~i\,g_s\int d^4x\int \frac{d^4k}{(2\pi)^4}\,e^{-i(k+p+q)\cdot x}\int_0^1 du \,\langle\, \rho^-(p,\eta^*)\,|\,\bar{d}(x)\, \gamma_{\mu \perp}\, \Big[\frac{u\,x_\alpha}{k^2-m_Q^2}\,\gamma_\beta \nonumber\\
&-\frac{\slashed{k}+m_Q}{2\,(k^2-m_Q^2)^2}\,\sigma_{\alpha\beta}\Big]\, G^{\alpha\beta}(ux)\,\gamma_5\,q(0)\,|\,0\rangle\,.
\label{p3}
\end{align}
Employing the conformal expansion of the three-particle DAs in appendix \ref{higher-das}, we obtain
\begin{align}
g^{{\rm3P}}=-\frac{m_Q}{f_H\,f_{H^*}\,m_H^2\,m_{H^*}}\,e^{\frac{m_H^2+m_{H^*}^2-2m_Q^2}{M^2}}\,f_\rho^T(\mu)\,m_\rho^2\, \Big[ \Big(\frac{21}{8}-4\ln2\Big)\,\tilde{t}_{10}(\mu)-\frac{15}{8}\,s_{00}(\mu)\Big]\,.\label{p3de2}
\end{align}

Collecting (\ref{twist2nlo}) (\ref{ht2p-result}) and (\ref{p3de2}), the final LCSR reads
\begin{align}
g_{H^*H\rho}=&-\frac{m_Q}{f_H\,f_{H^*}\,m_H^2\,m_{H^*}}\,e^{\frac{m_H^2+m_{H^*}^2-2m_Q^2}{M^2}}\,\bigg\{ f_\rho^T(\mu)\,\Big[ -\frac{M^2}{2}\,\Big(1-e^{-\frac{2s_0-2m_Q^2}{M^2}}\Big)+\frac{m_\rho^2}{4} \Big]\, \phi_\perp\Big(\frac{1}{2},\mu\Big)\nonumber\\
&+f_\rho^T(\mu)\,\mathcal{F}^{{\rm LT},(1)}\,e^{\frac{2m_Q^2}{M^2}} -\frac{m_Q}{2}\,m_\rho\,f_\rho\,g_\perp^{(a)}\Big(\frac{1}{2},\mu\Big) \nonumber\\
&+m_\rho^2\,f_\rho^T(\mu)\, \Big[\frac{1}{2}\, \Big(\frac{1}{2}+\frac{m_Q^2}{M^2}\Big)\, \mathbb{A}_T\Big(\frac{1}{2},\,\mu\Big)+ \Big(\frac{21}{8}-4\ln2\Big)\,\tilde{t}_{10}(\mu)-\frac{15}{8}\,s_{00}(\mu)\Big]\bigg\}\,,
\end{align}
where $\mathcal{F}^{{\rm LT},(1)}$ is defined in (\ref{nlo-ftw2}).

\section{Numerical analysis}
\label{sec-4}

\subsection{Input parameters}

The masses of quarks in $\overline{\rm MS}$ scheme and the values of decay constants are listed in Table \ref{decayconstant}, and values of the parameters in $\rho$ DAs  are listed in Table \ref{parameters-DAs}, which are extracted from the method of QCDSR \cite{Straub:2015ica,Ball:1998ff}.
\begin{table}[!htb]
\begin{center}
\begin{tabular}{c|c|c|c}
\hline
\hline
$\overline{m}_c(\overline{m}_c)$ (GeV) & $\overline{m_b}(\overline{m}_b)$ (GeV)  &   $f_\rho$ (MeV) & $f_\rho^T(\mu_0)$ (MeV) \\
\hline
$1.288\pm0.02$   & $4.193_{-0.035}^{+0.022}$  & $213\pm5$   & $160\pm7$ \\
\hline
$f_D$ (MeV)  & $f_{D^*}$ (MeV) & $f_B$ (MeV)   &  $f_{B^*}$ (MeV) \\
\hline
$209.0\pm2.4$  & $225.3\pm 8.0$  & $192.0\pm4.3$  & $182.4\pm 6.2$  \\
\hline
\hline
\end{tabular}
\end{center}
\caption{Values of heavy quark masses \cite{Dehnadi:2015fra,Beneke:2014pta} and decay constants \cite{Straub:2015ica,Aoki:2019cca,Lubicz:2017asp,Colquhoun:2015oha}, with the scale dependent quantity $f_\rho^T$ given at $\mu_0=1.0$~GeV.}
\label{decayconstant}
\end{table}

\begin{table}[!htb]
\begin{center}
\begin{tabular}{c|c|c|c|c|c}
\hline
\hline
$a_2^\parallel(\mu_0)$ & $a_2^\perp(\mu_0)$ & $\zeta_3(\mu_0)$ & $\omega_3^A(\mu_0)$ &$\omega_3^V(\mu_0)$ &$\omega_3^T(\mu_0) $\\
\hline
$0.17\pm0.07$ & $0.14\pm0.06$ & $0.032\pm0.010$ & $-2.1\pm1.0$& $3.8\pm1.8$  &$7.0 \pm 7.0$\\
\hline
$\zeta_4^T(\mu_0)$ & $\tilde{\zeta}_4^T(\mu_0)$ & $\langle\langle Q^{(1)}\rangle\rangle(\mu_0)$ & $\langle\langle Q^{(3)}\rangle\rangle(\mu_0)$ &  $\langle\langle Q^{(5)}\rangle\rangle(\mu_0)$ & \\
\hline
 $0.10\pm0.05$ & $-0.10 \pm 0.05$ &$-0.15\pm0.15$&$0$ & $0$  &  \\
\hline
\hline
\end{tabular}
\end{center}
\caption{Values of the nonperturbative parameters in the DAs  at the scale $\mu_0= 1.0 ~{\rm GeV}$.}
\label{parameters-DAs}
\end{table}

The solution to the two-loop evolution of the Gegenbauer moment $a_n^\perp(\mu)$ is
\begin{align}
f_\rho^T(\mu)\,a_n^\perp(\mu)=\Big[
E_{T, n}^{\rm NLO}(\mu, \mu_0)+\frac{\alpha_s(\mu)}{4\pi} \,\sum_{k=0}^{n-2}\, E_{T, n}^{\rm LO}(\mu,\mu_0)\,d_{T, n}^{k}(\mu,\mu_0) \Big]\,f_\rho^T(\mu_0)\,a_n^\perp(\mu_0)\,,
\end{align}
where the explicit expressions of $E_{T,n}$ and $d_{T,n}$ can be found in \cite{Wang:2017ijn}. The scale evolution of other nonperturbative parameters to leading logarithmic accuracy is listed in appendix \ref{higher-das}.

The factorization scales are taken as $\mu_c\in[1,\,2] $ GeV around the default choice $m_c$ and $\mu_b={m_b}^{+m_b}_{-m_b/2}$ for radiative $D^*$  and $B^*$ decays, respectively. As for the Borel mass $M^2$ and the threshold parameter $s_0$, we take the following intervals \cite{Khodjamirian:2009ys,Wang:2018wfj}
\begin{eqnarray}
&&\left\{
\begin{array}{l}
s_0=6.0\pm0.5 ~{\rm GeV}^2\,,\\
M^2=4.5\pm1.0 ~{\rm GeV}^2\, ,
\end{array}
\right. {\rm for} \,D^*\,D\,\rho\,;\quad
\left\{
\begin{array}{l}
s_0=34.0\pm1.0 ~{\rm GeV}^2\,,\\
M^2=18.0\pm3.0 ~{\rm GeV}^2\, ,
\end{array}
\right. {\rm for}\, B^*\,B\,\rho\,,
\end{eqnarray}
which satisfy the standard criterions \cite{Wang:2015vgv}.

\subsection{Theory predictions}
 The coupling constants for $D^*D\rho$ and $B^*B\rho$ are listed in Table \ref{results}. It's apparent that the perturbative QCD corrections decrease the  tree level leading power contributions of coupling constants by $10\%$ and $20\%$ for $D^*D\rho$ and $B^*B\rho$ respectively. Meanwhile, contributions from tree level sub-leading power corrections have obvious compensation effects, which increase the leading twist contributions by approximately $20\%$ and $30\%$ for $D^*D\rho$ and $B^*B\rho$ respectively. We also note that three-particle sub-leading power corrections have more minor effects compared with two-particle sub-leading power corrections, particular for coupling of $B^*B\rho$, which is consistent with the validity of OPE.

\begin{table}[!htb]
\begin{center}
\begin{tabular}{c|c|c|c|c|c}
\hline
\hline
& $g^{\rm LT,LL}$ & $g^{\rm LT,NLL}$ & $g^{\rm 2P,LL}$  & $g^{\rm 3P,LL}$ & Total   \\
\hline
$D^*\,D\,\rho$  & $3.61_{-0.47}^{+0.54}$  & $3.18_{-0.43}^{+0.53}$ & $0.47_{-0.16}^{+0.17}$ & $0.14_{-0.08}^{+0.09}$ & $3.80_{-0.45}^{+0.59}$ \\
\hline
$B^*\,B\,\rho$  & $3.59_{-0.43}^{+0.55}$  & $2.91_{-0.42}^{+0.38}$ & $0.96_{-0.15}^{+0.23}$ & $0.022_{-0.012}^{+0.013}$ & $3.89_{-0.48}^{+0.52}$ \\
\hline
\hline
\end{tabular}
\end{center}
\caption{The results  of $g_{H^*H\rho}$ ( ${\rm GeV}^{-1}$ ). Here  $g^{\rm LT, LL}$ corresponds to the tree-level leading power contributions with resummation at LL accuracy; $g^{\rm LT, NLL}$ corresponds to the leading power contributions up to NLO with resummation at NLL accuracy; $g^{\rm 2P, LL}$ and $g^{\rm 3P, LL}$ are from 2-particle and 3-particle sub-leading power corrections at LO separately.}
\label{results}
\end{table}

The uncertainties are also included for separate terms of coupling constants in Table \ref{results}. We mention that the uncertainties are estimated by varying independent input parameters, and the individual uncertainties are presented in Table \ref{individual-err}. To get the total uncertainties, we add them in quadrature to obtain the final results. From Table \ref{individual-err} we can see that the primary uncertainties are from Borel mass $M^2$ and parameter $a_2^\perp$ both for $D^*D\rho$ and $B^*B\rho$. The second Gegenbauer moment $a_2^\perp$ refers to transversely polarized $\rho$ meso twist-4 LCDA, which is indicated in appendix \ref{rhodas}, \ref{higher-das}. We mention that quasi distribution amplitudes (quasi-DAs), which are based on the large-momentum effective theory (LaMET) \cite{Ji:2013dva, Ji:2014gla}, together with the simulation from lattice QCD \cite{Braun:2016wnx} can also extract meson LCDA  \cite{Xu:2018mpf, Liu:2018tox} other than QCDSR. In the future, a more precise determination of the $\rho$ meson LCDA are helpful to decrease the uncertainties of the couplings.

\begin{table}[!htb]
\begin{center}
\begin{tabular}{l|c|c|c|c|c|c|c}
\hline
\hline
 & central value & $\Delta\, f_{H^*}$ & $\Delta\, m_{Q}$  &  $\Delta\, M^2$ & $\Delta\, \mu$ &$\Delta\, f_\rho^T$ & $\Delta\, a_2^\perp$ \\
\hline
$g_{D^*D\rho}$ & $3.80_{-0.45}^{+0.50}$  & $^{+0.14}_{-0.13}$ & $^{+0.00}_{-0.01}$  & $^{+0.38}_{-0.20}$ & $^{+0.20}_{-0.01}$ & $\pm0.11$ & $\pm 0.32$\\
\hline
$g_{B^*B\rho}$ & $3.89_{-0.48}^{+0.52}$  & $^{+0.14}_{-0.13}$ & $^{+0.12}_{-0.08}$  & $^{+0.36}_{-0.23}$ & $^{+0.14}_{-0.26}$ & $\pm0.12$ & $\pm 0.24$\\
\hline
\hline
\end{tabular}
\end{center}
\caption{The central value and individual uncertainty of $g_{H^*H\rho}$ ( $\rm{GeV}^{-1}$) due to the variations of input parameters. We only show the numerically significant uncertainties here. }
\label{individual-err}
\end{table}

 Now it's time to compare our results of coupling constants with others, we collect the results from several works which use sum rules as in Table \ref{others}. Among them, the LCSR works \cite{Li:2002pp, Wang:2007mc} didn't take the perturbative QCD corrections and 3-particle corrections into account, the QCDSR work considers the dimension-5 quark-gluon condensate corrections \cite{Khosravi:2014rwa}.   As mentioned before, the NLO effects of leading power almost cancel with the subleading-power corrections numerically, so it's natural that our results are close to previous sum rules works within errors. 

\begin{table}[!htb]
\begin{center}
\begin{tabular}{l|c|c|c|c}
\hline
\hline
 & This work & LCSR \cite{Li:2002pp} & LCSR \cite{Wang:2007mc}  & QCDSR \cite{Khosravi:2014rwa} \\
\hline
$g_{D^*D\rho}$ & $3.80_{-0.45}^{+0.50}$  & $4.17\pm1.04$ & $3.56\pm0.60$  & $4.07\pm0.71$\\
\hline
$g_{B^*B\rho}$ & $3.89_{-0.48}^{+0.52}$  & $5.70\pm1.43$ &  --   &--  \\
\hline
\hline
\end{tabular}
\end{center}
\caption{Numerical values of coupling constants $g_{H^*H\rho}$ ($\rm{GeV}^{-1}$) from several works via sum rules. }
\label{others}
\end{table}

Next we are going to extract the $B^*B\rho$ coupling from from factors. The $B\to\rho$ form factors $V(q^2)$ and $T_1(q^2)$ which relate $g_{B^*B\rho}$ are defined as
\begin{align}
&\langle\,\rho^-(p,\eta^*)\,|\,\bar{d}\,\gamma_\mu\,b\,|\,B^-(p+q)\,\rangle=\frac{2}{m_B+m_\rho}\,V(q^2)\,\epsilon_{\mu\eta pq}\,, \nonumber\\
&\langle\,\rho^-(p,\eta^*)\,|\,\bar{d}\,i\,\sigma_{\mu\nu}\,q^\nu\,b\,|\,B^-(p+q)\,\rangle=-2\,T_1(q^2)\,\epsilon_{\mu\eta pq}\,.
\end{align}
From the dispersion relation of form factor $F_i(q^2)$
\begin{align}
F_i(q^2)=\frac{r_1^i}{1-q^2/m_{B^*}^2}+\int_{(m_B+m_\rho)^2}^\infty \frac{\rho(s)}{s-q^2-i\epsilon}\,,
\end{align}
the strong coupling relates to the pole of the form factors at the unphysical point $q^2=m_{B^*}^2$. Then we have the relation
\begin{align}
r_1^V=&\lim_{q^2\to m_{B^*}^2} \big(1-q^2/m_{B^*}^2\big)\,V(q^2)=\frac{m_B+m_\rho}{2\,m_B^*}\,f_{B^*}\,g_{B^*B\rho}\,, \nonumber\\
r_1^{T_1}=&\lim_{q^2\to m_{B^*}^2} \big(1-q^2/m_{B^*}^2\big)\,T_1(q^2)=\frac{1}{2}\,f_{B^*}^T\,g_{B^*B\rho}\,,
\label{relation}
\end{align}
where $f_{B^*}^T$ is the tensor coupling of the $B^*$ meson and defined as
\begin{align}
\langle\,0\,|\,\bar{b}\,\sigma_{\mu\nu}\,q\,|\,B^*(q,\epsilon)\,\rangle=i\,f_{B^*}^T\,(\epsilon_\mu\,q_\nu-\epsilon_\nu\,q_\mu)\,.
\end{align}
Using (\ref{relation}) and choose the size $f_{B^*}^T=f_{B^*}$, we extract several  numerical values of the coupling $g_{B^*B\rho}$ from recent LCSR works \cite{Straub:2015ica,Gao:2019lta}, which are listed in table \ref{relation-ff}.
\begin{table}[!htb]
\begin{center}
\begin{tabular}{l|c|c|c|c}
\hline
\hline
 This work & $V(q^2)$ \cite{Straub:2015ica} & $T_1(q^2)$ \cite{Straub:2015ica}  &  $V(q^2)$ \cite{Gao:2019lta} &$T_1(q^2)$ \cite{Gao:2019lta} \\
\hline
$3.89_{-0.48}^{+0.52}$  & $8.50\pm1.73$ & $8.02\pm1.59$  & $6.04_{-2.34}^{+1.30}$ & $7.25^{+1.72}_{-2.78}$ \\
\hline
\hline
\end{tabular}
\end{center}
\caption{The coupling $g_{B^*B\rho}$ ($\rm{GeV}^{-1}$) from the residue of  $B\to\rho$ form factor $V$ and $T_1$ in LCSR fit and compared to our result.}
\label{relation-ff}
\end{table}


As it shows, our central value is smaller than the extrapolations from LCSR form factors. To explain this discrepancy, on one hand,  uncertainties from the parameterizations of the form factors to obtain the un-physical singularity may be underestimated. On the other hand, for the double dispersion relation, the Borel suppression is not sufficient enough for us to neglect the isolated excitation contributions following the discussion in \cite{Becirevic:2002vp}. Apart from $B \to \rho$, the $B^* \to B$ and $B^* \to \rho$ form factors can also be used to extract the coupling. To the best of our knowledge, we have not found relevant works that address corresponding form factors in LCSR. It's a future work to check that for the coupling extractions, whether different choices of form factors  are consistent.


At last, we analyze our work with other model-dependent works. The HM$\chi$PT effective Lagrangian to parametrize the $H^*HV$ coupling can be written as \cite{Casalbuoni:1996pg}
\begin{align}
\mathcal{L}_V=i\,\lambda\,{\rm Tr}[\mathcal{H}_b\,\sigma^{\mu\nu}\,F_{\mu\nu}(\rho)_{ba}\,\bar{\mathcal{H}}_a]\,,
\end{align}
with
\begin{align}
F_{\mu\nu}(\rho)=\partial_\mu\, \rho_\nu-\partial_\nu\, \rho_\mu +[\rho_\mu,\,\rho_\nu]\,, \quad \rho_\mu=i\,\frac{g_V}{\sqrt{2}}\,\hat{\rho}_\mu\,,
\end{align}
where $\hat{\rho}$ is $3\times 3$ matrix for light meson nonet, and the heavy $H, H^*$ mesons are represented by the doublet field $\mathcal{H}_a$ with the conventional normalization. The parameter $g_V= m_\rho/f_\pi$. In the chiral and heavy quark limits, we have the following relation for the coupling $\lambda$
\begin{align}
\lambda=\frac{\sqrt{2}}{4}\,\frac{1}{g_V}\,g_{H^*H\rho}\,.
\end{align}

 We find the coupling $\lambda=0.23\pm0.03\,{\rm GeV}^{-1}$ at leading power from our value of $g_{B^*B\rho}$.

\begin{table}[!htb]
\begin{center}
\begin{tabular}{c|c|c|c|c}
\hline
\hline
 This work &  VMD \cite{Isola:2003fh} & CQM \cite{Deandrea:1998ww} & CQM \cite{Melikhov:2000yu}   &QM+VMD \cite{Oh:2000qr} \\
\hline
$0.23$ & $0.56$  & $0.60$ & $0.47$   & $0.33$ \\
\hline
\hline
\end{tabular}
\end{center}
\caption{The central value of coupling $\lambda$ ($\rm{GeV}^{-1}$) from model estimations  compared to our result.}
\label{eff}
\end{table}

In table \ref{eff} we compare this value with other model estimations. Similar to the estimations from form factors, our result is smaller than model predictions.

One possibility for this discrepancy may be that the model predictions have potential larger errors. So the LCSR is more believable in principle. However, besides the possible influences of excitation contributions mentioned above, the values of NLO corrections to higher twists and the sub-sub-leading power contributions are unknown, which may also be sources of the discrepancy. Improvements for the calculations from both sum rules and models may help to understand such discrepancies in the future.


\section{Conclusion}
\label{sec-5}

We compute the $D^*D\rho$ and $B^*B\rho$ strong couplings to subleading-power in LCSR. The long-distance dynamics are incorporated in the $\rho$ DAs.  We calculated the $\mathcal{O}(\alpha_s)$ corrections to leading power of the sum rules. The subleading-power corrections are calculated at LO by accounting the two-particle and three-particle wave functions up to twist-4. The analytical results of double spectral densities are obtained, and with respect to the previous work, we also performed a continuum subtraction for the higher-twist corrections. The LO results are independent of the choice of duality region since the special form of the double spectral density.

Numerically, the NLO corrections decrease the tree-level results by $10\%$ and $20\%$ for $D^*D\rho$ and $B^*B\rho$ respectively. However, the subleading-power corrections at LO can give rise to the values about $20\%$ and $30\%$ for $D^*D\rho$ and $B^*B\rho$ respectively. Summing up all the contributions, our  values are consistent with the predictions from previous sum rules works. Moreover, we also predict the coupling $\lambda$ in HM$\chi$PT at leading power. The central value of our result is smaller than the existing model-dependent estimations. Possible explanations for the discrepancy refer to the potential large errors within models, and influences of excitation contributions in LCSR. Moreover, radiative corrections to higher twists and sub-sub-leading power contributions may also be sources for that. In the future, a better understanding for this discrepancy is beneficial to shed light on the long-distance QCD dynamics.


\section*{Acknowledgment}
We are grateful to Prof. Cai-dian L\"u and Prof. Yue-long Shen for helpful discussions. This work was supported in part by National Natural Science Foundation of China under Grant No. 11521505.


\appendix


\section{The $\rho$ meson DAs}
\label{rhodas}

Here we collect the $\rho$ meson DAs up to twist-4 which are defined in \cite{Ball:1998ff}
\begin{align}
&\langle\,\rho^-(p,\eta^*)\,|\,\bar{d}(x)\,\gamma_\mu\,u(0)\,|\,0\,\rangle \nonumber \\
=&~f_\rho\,m_\rho\,\Big\{\frac{\eta^*\cdot x}{p\cdot x}\,p_\mu\int_0^1du\,e^{i\,u\,p\cdot x}\,\big[\phi_\parallel(u)+\frac{m_\rho^2\,x^2}{16}\,\mathbb{A}(u)\big] \nonumber\\
&+\eta^*_{\perp\mu}\int_0^1du\,e^{i\,u\,p\cdot x}\,g_\perp^{(v)}(u)-\frac{1}{2}\,x_\mu\,\frac{\eta^*\cdot x}{(p\cdot x)^2}\,m_\rho^2 \int_0^1du\,e^{i\,u\,p\cdot x}\,g_3(u)\Big\}\,, \\
&\langle\,\rho^-(p,\eta^*)\,|\,\bar{d}(x)\,\gamma_\mu\,\gamma_5\,u(0)\,|\,0\,\rangle=\frac{1}{4}\,f_\rho\,m_\rho\,\epsilon_{\mu\eta px} \int_0^1du\,e^{i\,u\,p\cdot x}\,g_\perp^{(a)}(u)\,, \\
&\langle\,\rho^-(p,\eta^*)\,|\,\bar{d}(x)\,\sigma_{\mu\nu}\,u(0)\,|\,0\,\rangle \nonumber\\
=&-i\,f_\rho^T\,\bigg\{ (\eta^*_{\perp\mu}\,p_\nu-\eta^*_{\perp\nu}\,p_\mu) \int_0^1du\,e^{i\,u\,p\cdot x}\, \big[\phi_\perp(u)+\frac{m_\rho^2\,x^2}{16}\,\mathbb{A}_T(u)\big] \nonumber\\
&+(p_\mu\,x_\nu-p_\nu\,x_\mu)\,\frac{\eta^*\cdot x}{(p\cdot x)^2}\,m_\rho^2 \int_0^1du\,e^{i\,u\,p\cdot x}\,h_\parallel^{(t)}(u) \nonumber\\
&+\frac{1}{2}\,\frac{\eta^*_{\perp\mu}\,x_\nu-\eta^*_{\perp\nu}\,x_\mu}{p\cdot x}\,m_\rho^2 \int_0^1du\,e^{i\,u\,p\cdot x}\,h_3(u) \bigg\}\,, \\
&\langle\,\,\rho^-(p,\eta^*)\,|\,\bar{d}(x)\,u(0)\,|\,0\,\rangle=-\frac{i}{2}\,f_\rho^T\,(\eta^*\cdot x) \,m_\rho^2 \int_0^1du\,e^{i\,u\,p\cdot x}\,h_\parallel^{(s)}(u)\,, \\
&\langle\,\rho^-(p,\eta^*)\,|\,\bar{d}(x)\,g_s\,\tilde{G}_{\mu\nu}(vx)\,\gamma_\alpha\,\gamma_5\,u(0)\,|\,0\,\rangle \nonumber\\
=&-f_\rho\,m_\rho\,p_\alpha\,\big[p_\nu\, \eta^*_{\perp\mu}-p_\mu\,\eta^*_{\perp\nu}\big]\int[D\alpha]\,e^{i\,(\alpha_q+v\,\alpha_g)\,p\cdot x}\,\mathcal{A}(\alpha) \nonumber\\
&-f_\rho\,m_\rho^3\,\frac{\eta^*\cdot x}{p\cdot x}\,\big[p_\mu\,g^\perp_{\alpha\nu}-p_\nu\,g^\perp_{\alpha\mu}\big] \int[D\alpha]\,e^{i\,(\alpha_q+v\,\alpha_g)\,p\cdot x}\,\tilde{\Phi}(\alpha)\nonumber\\
&-f_\rho\,m_\rho^3\,\frac{\eta^*\cdot x}{(p\cdot x)^2}\, p_\alpha\, \big[p_\mu\,x_\nu-p_\nu\,x_\mu\big]\int[D\alpha]\,e^{i\,(\alpha_q+v\,\alpha_g)\,p\cdot x}\, \tilde{\Psi}(\alpha)\,, \label{dasa5}\\
&\langle\,\rho^-(p,\eta^*)\,|\,\bar{d}(x)\,g_s\,G_{\mu\nu}(vx)\,i\,\gamma_\alpha\,u(0)\,|\,0\,\rangle \nonumber\\
=&-f_\rho\,m_\rho\,p_\alpha\,\big[p_\nu\, \eta^*_{\perp\mu}-p_\mu\,\eta^*_{\perp\nu}\big]\int[D\alpha]\,e^{i\,(\alpha_q+v\,\alpha_g)\,p\cdot x}\,\mathcal{V}(\alpha) \nonumber\\
&-f_\rho\,m_\rho^3\,\frac{\eta^*\cdot x}{p\cdot x}\,\big[p_\mu\,g^\perp_{\alpha\nu}-p_\nu\,g^\perp_{\alpha\mu}\big] \int[D\alpha]\,e^{i\,(\alpha_q+v\,\alpha_g)\,p\cdot x}\,\Phi(\alpha)\nonumber\\
&-f_\rho\,m_\rho^3\,\frac{\eta^*\cdot x}{(p\cdot x)^2}\, p_\alpha\, \big[p_\mu\,x_\nu-p_\nu\,x_\mu\big]\int[D\alpha]\,e^{i\,(\alpha_q+v\,\alpha_g)\,p\cdot x}\,\Psi(\alpha)\,, \\
&\langle\,\rho^-(p,\eta^*)\,|\,\bar{d}(x)\,g_s\,G_{\mu\nu}(vx)\,\sigma_{\alpha\beta}\,u(0)\,|\,0\,\rangle \nonumber\\
=&~f^T_\rho\,m_\rho^2\,\frac{\eta^*\cdot x}{2\,p\cdot x} \,\big[ p_\alpha\,p_\mu\,g^\perp_{\beta\nu}-p_\beta\,p_\mu\,g^\perp_{\alpha\nu}-(\mu\leftrightarrow \nu) \big] \int[D\alpha]\,e^{i\,(\alpha_q+v\,\alpha_g)\,p\cdot x}\,\mathcal{T}(\alpha)\nonumber\\
&+f^T_\rho\,m_\rho^2\,\big[ p_\alpha\,\eta^*_{\perp\mu}\,g^\perp_{\beta\nu}-p_\beta\,\eta^*_{\perp\mu} \,g^\perp_{\alpha\nu}-(\mu\leftrightarrow \nu) \big] \int[D\alpha]\,e^{i\,(\alpha_q+v\,\alpha_g)\,p\cdot x}\,T_1^{(4)}(\alpha)\nonumber\\
&+f^T_\rho\,m_\rho^2\,\big[ p_\mu\,\eta^*_{\perp\alpha}\,g^\perp_{\beta\nu}-p_\mu\,\eta^*_{\perp\beta} \,g^\perp_{\alpha\nu} -(\mu\leftrightarrow \nu) \big] \int[D\alpha]\,e^{i\,(\alpha_q+v\,\alpha_g)\,p\cdot x}\,T_2^{(4)}(\alpha)\nonumber\\
&+f^T_\rho\,m_\rho^2\,\frac{(p_\mu\,x_\nu-p_\nu\,x_\mu)\,(p_\alpha\,\eta^*_{\perp\beta}-p_\beta\,\eta^*_{\perp\alpha})}{p\cdot x} \int[D\alpha]\,e^{i\,(\alpha_q+v\,\alpha_g)\,p\cdot x}\,T_3^{(4)}(\alpha)\nonumber\\
&+f^T_\rho\,m_\rho^2\,\frac{(p_\alpha\,x_\beta-p_\beta\,x_\alpha)\,(p_\mu\,\eta^*_{\perp\nu}-p_\nu\,\eta^*_{\perp\mu})}{p\cdot x} \int[D\alpha]\,e^{i\,(\alpha_q+v\,\alpha_g)\,p\cdot x}\,T_4^{(4)}(\alpha)\,,\\
&\langle\,\rho^-(p,\eta^*)\,|\,\bar{d}(x)\,g_s\,G_{\mu\nu}(vx)\,u(0)\,|\,0\,\rangle \nonumber\\
=&-i\,f^T_\rho\,m_\rho^2\,\big[\eta^*_{\perp\mu}\,p_\nu-\eta^*_{\perp\nu}\,p_\mu\big]\int[D\alpha]\,e^{i\,(\alpha_q+v\,\alpha_g)\,p\cdot x}\,S(\alpha)\,,\\
&\langle\,\rho^-(p,\eta^*)\,|\,\bar{d}(x)\,i\,g_s\,\tilde{G}_{\mu\nu}(vx)\,\gamma_5\,u(0)\,|\,0\,\rangle \nonumber\\
=&~i\,f^T_\rho\,m_\rho^2\,\big[\eta^*_{\perp\mu}\,p_\nu-\eta^*_{\perp\nu}\,p_\mu\big]\int[D\alpha]\,e^{i\,(\alpha_q+v\,\alpha_g)\,p\cdot x}\,\tilde{S}(\alpha)\,, \label{das5}
\end{align}
where
\begin{align}
&\widetilde{G}_{\alpha \beta}= \frac{1}{2} \,  \epsilon_{\alpha \beta \rho \tau } \, G^{\rho \tau} \,.
\end{align}


\section{The conformal expansion of $\rho$ DAs}
\label{higher-das}
Here we list the conformal expansion for these higher twist DAs involved in our calculation \cite{Ball:1998ff}.

The chiral-even twist-three DA $g_\perp^{(a)}$ reads
\begin{align}
g_\perp^{(a)}(u)=6\,u\,\bar{u}\,\Big\{1+ \Big[\frac{1}{4}\,a_2^\parallel +\frac{5}{3}\,\zeta_3\,\Big(1-\frac{3}{16}\,\omega_3^A +\frac{9}{16}\,\omega_3^V\Big)\Big] \,(5\,\xi^2-1)  \Big\}\,,
\end{align}
with $\xi = 2\,u-1$. For the chiral-odd twist-four DA $\mathbb{A}_T$ we have
\begin{align}
\mathbb{A}_T(u)=&~30\,u^2\,\bar{u}^2\,\Big[\frac{2}{5}\,\Big(1+\frac{2}{7}\,a_2^\perp+ \frac{10}{3}\,\zeta_4^T- \frac{20}{3}\,\tilde{\zeta}_4^T\Big)+\Big(\frac{3}{35}\,a_2^\perp+\frac{1}{40}\,\zeta_3\,\omega_3^T\Big)\,C_2^{5/2}(\xi)\Big]  \nonumber\\
&-\Big[\frac{18}{11}\,a_2^\perp-\frac{3}{2}\,\zeta_3\,\omega_3^T+ \frac{126}{55}\,\langle\langle Q^{(1)}\rangle\rangle +\frac{70}{11}\,\langle\langle Q^{(3)}\rangle\rangle\Big]\, \big[u\,\bar{u}\,(2+13\,u\,\bar{u}) \nonumber\\
&+2\,u^3\,(10-15\,u+6\,u^2)\,\ln u +2\,\bar{u}^3\,(10-15\,\bar{u}+6\,\bar{u}^2)\,\ln \bar{u}\big]\,.
\end{align}
As for the twist-4 three-particle DAs
\begin{align}
S(\alpha_i)=&~30\,\alpha_g^2\,\Big\{ s_{00}\,(1-\alpha_g)+s_{10}\, \Big[ \alpha_g\,(1-\alpha_g)-\frac{3}{2}\,(\alpha_{\bar{q}}^2+ \alpha_q^2) \Big] \nonumber\\
&+s_{01}\, \big[ \alpha_g\,(1-\alpha_g)-6\,\alpha_{\bar{q}}\,\alpha_q \big]\Big\}\,, \nonumber\\
\tilde{S}(\alpha_i)=&~30\,\alpha_g^2\,\Big\{ \tilde{s}_{00}\,(1-\alpha_g)+\tilde{s}_{10}\, \Big[ \alpha_g\,(1-\alpha_g)-\frac{3}{2}\,(\alpha_{\bar{q}}^2+ \alpha_q^2) \Big] \nonumber\\
&+\tilde{s}_{01}\, \big[ \alpha_g\,(1-\alpha_g)-6\,\alpha_{\bar{q}}\,\alpha_q \big]\Big\}\,, \nonumber\\
T_1^{(4)}(\alpha_i)=&~120\,t_{10}\,(\alpha_{\bar{q}}-\alpha_q)\, \alpha_{\bar{q}}\,\alpha_q\,\alpha_g\,, \nonumber\\
T_2^{(4)}(\alpha_i)=&-30\,\alpha_g^2\,(\alpha_{\bar{q}}-\alpha_q)\, \Big[\tilde{s}_{00}+\frac{1}{2}\, \tilde{s}_{10}\, (5\,\alpha_g-3)+\tilde{s}_{01}\,\alpha_g\Big]\,, \nonumber\\
T_3^{(4)}(\alpha_i)=&-120\,\tilde{t}_{10}\,(\alpha_{\bar{q}}-\alpha_q)\, \alpha_{\bar{q}}\,\alpha_q\,\alpha_g\,, \nonumber\\
T_4^{(4)}(\alpha_i)=&~30\,\alpha_g^2\,(\alpha_{\bar{q}}-\alpha_q)\, \Big[s_{00}+\frac{1}{2}\, s_{10}\, (5\,\alpha_g-3) +s_{01}\,\alpha_g\Big]\,.
\end{align}
The eight parameters in three-particle DAs can be written as
\begin{align}
s_{00}&=\zeta_4^T\,\quad \tilde{s}_{00}=\tilde{\zeta}_4^T\,, \nonumber\\
s_{10}&=-\frac{3}{22}\,a_2^\perp-\frac{1}{8}\,\zeta_3\,\omega_3^T+\frac{28}{55}\,\langle\langle Q^{(1)}\rangle\rangle +\frac{7}{11}\,\langle\langle Q^{(3)}\rangle\rangle+ \frac{14}{3}\,\langle\langle Q^{(5)}\rangle\rangle\,, \nonumber\\
\tilde{s}_{10}&=\frac{3}{22}\,a_2^\perp-\frac{1}{8}\,\zeta_3\,\omega_3^T-\frac{28}{55}\,\langle\langle Q^{(1)}\rangle\rangle -\frac{7}{11}\,\langle\langle Q^{(3)}\rangle\rangle+ \frac{14}{3}\,\langle\langle Q^{(5)}\rangle\rangle\,, \nonumber\\
s_{01}&=\frac{3}{44}\,a_2^\perp+\frac{1}{8}\,\zeta_3\,\omega_3^T+\frac{49}{110}\,\langle\langle Q^{(1)}\rangle\rangle -\frac{7}{22}\,\langle\langle Q^{(3)}\rangle\rangle+ \frac{7}{3}\,\langle\langle Q^{(5)}\rangle\rangle\,, \nonumber\\
\tilde{s}_{01}&=-\frac{3}{44}\,a_2^\perp+\frac{1}{8}\,\zeta_3\,\omega_3^T-\frac{49}{110}\,\langle\langle Q^{(1)}\rangle\rangle +\frac{7}{22}\,\langle\langle Q^{(3)}\rangle\rangle+ \frac{7}{3}\,\langle\langle Q^{(5)}\rangle\rangle\,, \nonumber\\
t_{10}&=-\frac{9}{44}\,a_2^\perp-\frac{3}{16}\,\zeta_3\,\omega_3^T-\frac{63}{220}\,\langle\langle Q^{(1)}\rangle\rangle +\frac{119}{44}\,\langle\langle Q^{(3)}\rangle\rangle\,, \nonumber\\
\tilde{t}_{10}&=\frac{9}{44}\,a_2^\perp-\frac{3}{16}\,\zeta_3\,\omega_3^T+\frac{63}{220}\,\langle\langle Q^{(1)}\rangle\rangle +\frac{35}{44}\,\langle\langle Q^{(3)}\rangle\rangle\,.
\end{align}
The scale evolution of the nonperturbative parameters  to leading logarithmic accuracy is as follows
\begin{eqnarray}
&a_2^\parallel(\mu)=L^{\frac{25}{6}C_F/\beta_0}\,a_2^\parallel(\mu_0)\,,\quad
&\zeta_3(\mu)= L^{\left(-\frac{1}{3}C_F+3C_A\right) / \beta_0} \,\zeta_3(\mu_0)\,, \nonumber\\
&\omega_3^T(\mu)= L^{ \left(\frac{25}{6}C_F-2C_A\right) / \beta_0} \,\omega_3^T(\mu_0) \,,
&(\zeta_4^T+\tilde{\zeta}_4^T)(\mu)= L^{\left(3C_A-\frac{8}{3}C_F\right)/\beta_0}\,(\zeta_4^T+\tilde{\zeta}_4^T)(\mu_0)\,, \nonumber \\
&(\zeta_4^T-\tilde{\zeta}_4^T)(\mu)= L^{\left(4C_A-4C_F\right)/\beta_0}\,(\zeta_4^T-\tilde{\zeta}_4^T)(\mu_0)\,,
& \langle\langle Q^{(1)}\rangle\rangle(\mu)= L^{\left(-4C_F+\frac{11}{2}C_A\right)/\beta_0} \,\langle\langle Q^{(1)}\rangle\rangle(\mu_0)\,,  \nonumber \\
&\langle\langle Q^{(3)}\rangle\rangle(\mu)= L^{\frac{10}{3}C_F/\beta_0} \,\langle\langle Q^{(3)}\rangle\rangle(\mu_0)\,,
&\langle\langle Q^{(5)}\rangle\rangle(\mu)= L^{\left(-\frac{5}{3}C_F+5C_A\right)/\beta_0} \,\langle\langle Q^{(5)}\rangle\rangle(\mu_0)\,, \nonumber \\
\end{eqnarray}
where $L=\alpha_s(\mu)/\alpha_s(\mu_0)$ and $\beta_0=11-2\,n_f/3$, $n_f$ being the number of flavors involved. The scale-dependence of $\omega^{V(A)}_\gamma(\mu)$
\begin{eqnarray}
\left(
\begin{array}{c}
\omega^{V}_3(\mu) - \omega^{A}_3(\mu) \\
\omega^{V}_3(\mu) + \omega^{A}_3(\mu)
\end{array}
\right)
=L^{\Gamma_{\omega}  / \beta_0} \,
\left(
\begin{array}{c}
\omega^{V}_3(\mu_0) - \omega^{A}_3(\mu_0) \\
\omega^{V}_3(\mu_0) + \omega^{A}_3(\mu_0)
\end{array}
\right) \,,
\end{eqnarray}
where $\Gamma_{\omega}$ is given by
\begin{eqnarray}
\Gamma_{\omega}=
\left(
\begin{array}{c}
3 \, C_F - {2 \over 3} \, C_A  \qquad   {2 \over 3} \, C_F - {2 \over 3} \, C_A   \\
{5 \over 3} \, C_F - {4 \over 3} \, C_A \qquad  {1 \over 2} \, C_F  + C_A
\end{array}
\right)\,.
\end{eqnarray}


\end{document}